# Inertial Extended-Lagrangian Scheme for Solving Charge Equilibration Models

Itai Leven[1,2,5] and Teresa Head-Gordon[1,2,3,4,5]

[1]Kenneth S. Pitzer Center for Theoretical Chemistry, [2]Department of Chemistry, [3]Department of Bioengineering, [4]Department of Chemical and Biomolecular Engineering
University of California Berkeley
[5]Chemical Sciences Division, Lawrence Berkeley National Laboratory
Berkeley, California 94720, USA

The inertial extended Lagrangian/self-consistent field scheme (iEL-SCF) has been adopted for solving charge equilibration in LAMMPS as part of the reactive force field ReaxFF, which due to the charge conservation constraint requires solving two sets of linear system of equations for the new charges at each molecular dynamics time-step. Therefore, the extended Lagrangian for charge equilibration is comprised of two auxiliary variables for the intermediate charges which serve as an initial guess for the real charges. We show that the iEL-SCF is able to reduce the number of SCF cycles by 50-80% of the original conjugate gradient self-consistent field solver as tested across diverse systems including water, ferric hydroxide, nitramine RDX, and hexanitrostilbene.

[†]Corresponding author: thg@berkeley.edu

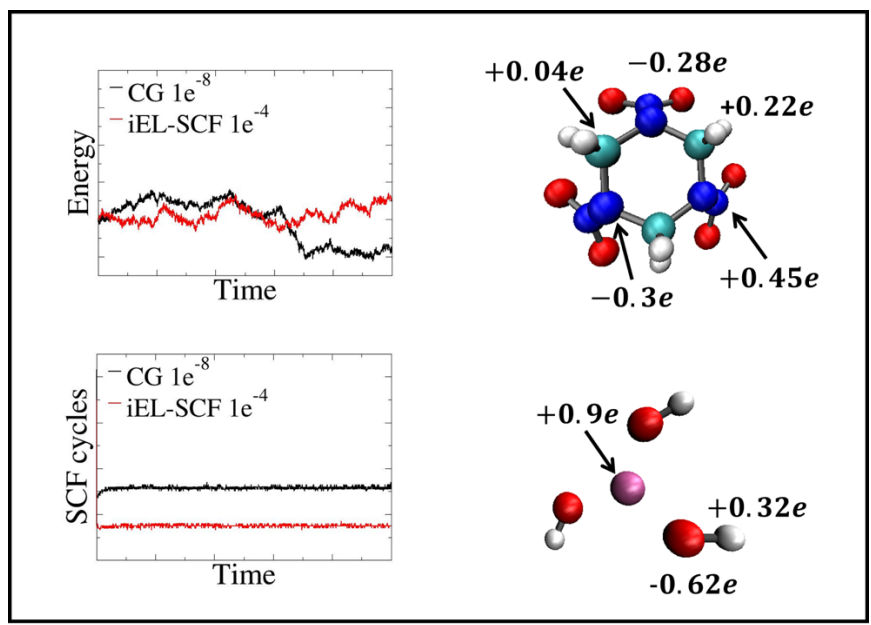

## INTRODUCTION

Charge equilibration force fields are many-body potentials for electronic charge rearrangements in molecules[1, 2], which have been widely used in molecular dynamics (MD) simulation of biological membranes and membrane proteins[3], nanoporous materials[4, 5], and are indispensable for accounting for the variation of charges in chemical reactions when using reactive force fields such as ReaxFF[6-8]. Starting with the concepts of atomic hardness and electronegativity in the framework of density functional theory (DFT) introduced by Parr and Pearson[9], Mortier *et al.* developed the electronegativity equalization method (EEM) to realize the charge rearrangements by adopting atomic electronegativity and hardness as fitting parameters to DFT derived Mulliken charges[1, 10]. Rappé and Goddard extended the EEM to what is today termed the charge equilibration method (CEM) by replacing the standard Coulomb potential with a shielded electrostatic term, and using experimental atomic ionization potentials, electron affinities, and atomic radii as the input data for optimizing the charge rearrangements in response to nuclear displacements.[2]

The basic CEM is the process of solving the linear system of equations for the new charges under a constraint that the net charge of the entire system is conserved, usually through iterative approaches such as conjugate gradient self-consistent field (CG-SCF) methods. In order to minimize the computational cost of the CG-SCF step, the use of well formulated preconditioners and initial guess extrapolations to improve convergence, as well as good software implementations on many-core hardware architectures, have been helpful[11-14]. Nonetheless, the solution of the many-body charge equilibration forces at each time step remains the most computationally demanding component of MD simulations using the many-body potential, especially using reactive potentials such as ReaxFF which are approximately tens to hundred times slower than traditional non-reactive force fields.

An alternative approach for solving a many-body potential is to instead dynamically propagate a set of auxiliary electronic variables using an extended Lagrangian (EL) formulation[15, 16] However, early invocations of EL schemes required an auxiliary mass parameter, $m$, that determined the trade-off between a small MD time step that can match the SCF solution accuracy vs. a desirable longer MD time-step with diminished ability to stay on the Born-Oppenheimer surface. A particularly elegant solution was proposed by Niklasson and co-workers, whereby in the limit that $m \to 0$ the auxiliary electronic degrees evolve time-reversibly under a harmonic potential that stays close to the Born-Oppenheimer surface by serving as an initial guess for an SCF solver[17, 18]. The resulting XL-BOMD method uses a regular MD time step, i.e. now determined by the specifics of the numerical integrator such as Verlet, and better maintains energy conservation with looser SCF convergence.[19]

However, the accumulated numerical error of the incomplete SCF convergence ultimately manifests in numerical instabilities and corrupted dynamics of the auxiliary electronic degrees of freedom, which in turn creates an unbounded increase in the number of SCF cycles required to achieve convergence[20, 21]. In fact, the XL-BOMD has been previously applied for solving charge equilibration[22], however the lack of dissipation to the auxiliary charges renders the method unstable at time scales exceeding even just a few picoseconds as we show later in the results. Niklasson *et al.* introduced various practical dissipation schemes, although they have the drawback of destroying time reversibility (albeit at very high order in the integration)[23]. Our group instead introduced an extended system that includes thermostats for the auxiliary electronic degrees of freedom that maintained time-reversibility[21]. The resulting inertial extended Lagrangian (iEL-SCF) has been shown to provide superior energy conservation at loose SCF convergence tolerance with reduced number of SCF cycles while remaining stable, and it has been used for molecular dynamics of many-body polarization potentials[21] and linear scaling forces based on DFT[24].

In this paper we have implemented the iEL-SCF method for solving charge equilibration in the framework of the ReaxFF force field in the LAMMPS molecular dynamics simulation platform[25]. Unlike the case of induced polarization, the additional constraint that the net charge of the entire system is conserved comes at the cost of having to solve two sets of linear equations for the intermediate charges $q^s$ and $q^t$, which are then combined to evaluate the real charges $q$, at each time step. We show that application of the iEL-SCF method to charge equilibration allows stable simulations on long timescales while reducing the number of SCF cycles to approximately half the number when compared to a CG-SCF solver using default settings in LAMMPS. The iEL/SCF for charge equilibration is shown to work well for various reactive systems from liquids to solids and at the high temperatures used in combustion applications.

**THEORY**

The CEM method is based on the second order Taylor expansion of the atomic energy with respect to partial charge around the (usually neutral) charge reference point $E_i(0)$:

$$E_i(q_i) = E_i(0) + q_i \left(\frac{\partial E_i}{\partial q_i}\right)_{q=0} + \frac{1}{2}q_i^2 \left(\frac{\partial^2 E_i}{\partial q_i^2}\right)_{q=0} \tag{1}$$

where $E_i(q)$ is the energy of atom $i$ given its atomic charge $q_i$, and $\left(\frac{\partial E_i}{\partial Q_i}\right)$ is the atomic electronegativity $\chi_i^0$ and $\frac{\partial^2 E_i}{\partial Q_i^2}$ is the atomic hardness $J_{ii}^0$ as motivated by Parr and Pearson[9]. Mortiel et al. subsequently introduced a Coulombic interaction between the charged atoms with the functional form[1, 10]:

$$E(q_1, q_2, \ldots, q_N) = \sum_{i=1}^{N} \left( E_i(0) + \chi_i^0 q_i + \frac{1}{2} \sum_{j=1}^{N} H_{ij} q_i q_j \right) \tag{2a}$$

where N is the number of atoms in the system and $H_{ij} = J_{ii}^0 \delta_{ij} + (1 - \delta_{ij})/r_{ij}$ where $\delta_{ij}$ is the Kronecker delta function and $r_{ij}$ is the distance between atoms $i$ and $j$. Subsequently Rappe and Goddard[2] reformulated $H_{ij}$ by replacing the bare Coulomb potential with a shielded electrostatic potential interaction term

$$H_{ij} = J_{ii}^0 \delta_{ij} + J_{ij}(1 - \delta_{ij}) \quad \text{where} \quad J_{ij} = \frac{1.0}{\left[ r_{ij}^3 + \left(\frac{1}{\gamma_{ij}}\right)^3 \right]^{\frac{1}{3}}} \tag{2b}$$

where $\gamma_{ij}$ an electrostatic shielding parameter. The resulting charge equilibration method postulates that the electronegativity of all the atoms in a molecule must equalize $\chi_1 = \chi_2 = \cdots = \chi_N$, which is determined by differentiating Eq. (2) with respect to the atomic charge:

$$\chi_i(q_1, q_2, \ldots, q_N) = \frac{\partial E}{\partial q_i} = \chi_i^0 + J_{ii}^0 q_i + \sum_{j \neq i} J_{ij} q_j \tag{3a}$$

while enforcing the constraint of charge neutrality:

$$\sum_{i=1}^{N} q_i = 0 \tag{3b}$$

In practice the new charges are determined by minimizing the system charge energy under the charge neutrality constraint using a Lagrange multiplier $\mu$:

$$E_\mu = E(q_1, q_2, \ldots, q_N) - \mu \left( \sum_{i=1}^{N} q_i - 0 \right) \tag{4a}$$

and by setting the derivative to zero:

$$\frac{\partial E_\mu}{\partial q_i} = \chi_i^0 + J_{ii}^0 q_i + \sum_{j \neq i} J_{ij} q_j - \mu = 0 \tag{4b}$$

It is evident that $\mu=\chi_i(q_1, q_2, \ldots, q_N)$ satisfies the electronegativity equalization condition. This allows us to formulate the charges in terms of $\mu$

$$q_i = \sum_{j=1}^{N} H_{ij}^{-1}(-\chi_j^0 + \mu \cdot 1_j) \tag{5a}$$

where the solution for $\mu$ is formulated from the constraint

$$\sum_{i=1}^{N} q_i = \sum_{i=1}^{N}\sum_{j=1}^{N} H_{ij}^{-1}(-\chi_j^0 + \mu \cdot 1_j) = 0 \tag{5b}$$

and is expressed as:

$$\mu = \frac{\sum_{i=1}^{N}\sum_{j=1}^{N} H_{ij}^{-1}(-\chi_j^0)}{\sum_{i=1}^{N}\sum_{j=1}^{N} H_{ij}^{-1}(1_j)} \tag{5c}$$

Eq. (5) is solved in LAMMPs by defining the numerator and denominator in Eq. (5c) in terms of two new variables, $q_s$ and $q_t$

$$\mu = \frac{\sum_{i=1}^{N}\sum_{j=1}^{N} H_{ij}^{-1}(-\chi_j^0)}{\sum_{i=1}^{N}\sum_{j=1}^{N} H_{ij}^{-1}(1_j)} = \frac{\sum_{i=1}^{N} q_i^s}{\sum_{i=1}^{N} q_i^t} \tag{6a}$$

to yield two systems of linear equations which are solved separately:

$$\sum_{i=1}^{N}\sum_{j=1}^{N} H_{ij}(q_i^s) = \sum_{i=1}^{N} -\chi_i^0 \tag{6b}$$

$$\sum_{i=1}^{N}\sum_{j=1}^{N} H_{ij}(q_i^t) = \sum_{i=1}^{N} 1_i \tag{6c}$$

to define the final partial charge of atom $i$ self-consistently as

$$q_i = q_{i,SCF}^s + \mu \cdot q_{i,SCF}^t \tag{6d}$$

In LAMMPS, Eq. (6) is solved using the CG-SCF method with a diagonal inverse baseline preconditioner and quadratic extrapolation of previous time-steps for the initial guess.

In this work we instead solve the two sets of linear equations for charge equilibration using the iEL-SCF method[21, 26-29] by formulating an extended Lagrangian with two additional auxiliary variables $q_{aux}^s$ and $q_{aux}^t$

$$\mathcal{L}_{hybrid}^{charge} = \frac{1}{2}\sum_{i=1}^{N} m_i \dot{\vec{r}}_i^2 + \frac{1}{2}\sum_{i=1}^{N} m_s (\dot{q}_{i,aux}^s)^2 + \frac{1}{2}\sum_{i=1}^{N} m_t (\dot{q}_{i,aux}^t)^2 - U(\vec{r}^N, q^N)$$

$$- \frac{1}{2}\omega^2 \sum_{i=1}^{N} m_s (q_{i,aux}^s - q_{i,SCF}^s)^2 - \frac{1}{2}\omega^2 \sum_{i=1}^{N} m_t (q_{i,aux}^t - q_{i,SCF}^t)^2$$

where $\vec{r}_i$ is the position vector of atom $i$, $U$ is the ReaxFF potential energy, the auxiliary variables evolve in time subject to a harmonic potential around $q_{i,SCF}^s$ and $q_{i,SCF}^t$, and $m_i$, $m_s$ and $m_t$ are the masses of the atom $i$, $q_{i,aux}^s$, and $q_{i,aux}^t$ auxiliary variables respectively. In the limit of $m_t \to 0$ and $m_s \to 0$ we recover the usual expression for the real degrees of freedom

$$m_i \ddot{\vec{r}}_i = -\frac{dU(\vec{r}^N, q^N)}{d\vec{r}_i} \tag{8}$$

and the corresponding equations of motion of the variables $s$ and $t$ are given by:

$$\ddot{q}_i^s = \omega^2 (q_{i,aux}^s - q_{i,SCF}^s) \qquad \ddot{q}_i^t = \omega^2 (q_{i,aux}^t - q_{i,SCF}^t) \tag{9}$$

At each time step the variables $q_{aux}^s$ and $q_{aux}^t$ are propagated together with the atomic degrees of freedom with a time reversible velocity Verlet integrator using a standard time step, $\Delta t$, which defines $\omega = \sqrt{2}/\Delta t$. Because the role of the variables $q_{aux}^s$ and $q_{aux}^t$ are to serve as quality initial guesses for the CG-SCF solutions, we have no need for the diagonal preconditioner of the original CG-SCF solver used in LAMMPS.

As diagnosed previously[21], the above solution of the extended Lagrangian yields unstable dynamics due to the inability to dissipate numerical error as a result of insufficient convergence of the CG-SCF, which manifests as corrupt dynamics in the auxiliary equations of motion. To circumvent this problem the iEL/SCF method is then defined by creating an inertial constraint on the auxiliary velocities in the form of thermostats. We have shown that the performance of Berendesen velocity rescaling is largely equivalent to that of using a Nose Hoover thermostat since the auxiliary variables are only initial guesses to the CG-SCF solution (unlike the case of our iEL/0-SCF method[27-30]). The Berendsen rescaling factor $\alpha$ of the auxiliary variables $q_{aux}^s$ and $q_{aux}^t$ is defined by:

$$\alpha_s = \sqrt{1 + \frac{\Delta t}{\tau}\left(\frac{T_s}{\langle(\dot{q}_{aux}^s)^2\rangle} - 1\right)} \qquad \alpha_t = \sqrt{1 + \frac{\Delta t}{\tau}\left(\frac{T_t}{\langle(\dot{q}_{aux}^t)^2\rangle} - 1\right)} \tag{10}$$

where $\tau$ is the rescaling parameter, $T_s$ and $T_t$ are the thermostat temperatures for $q_{aux}^s$ and $q_{aux}^t$ respectively, and the squared velocities are averaged over the ensemble.

## METHODS

The simulations were performed utilizing the ReaxFF force field[6-8] within the LAMMPS simulation package[25], for four different chemical systems: 140 $FeOH_3$ molecules and using the iron-oxyhydroxide force field[31], RDX $[CH_2N(NO_2)]_3$ system comprising 60 molecules and using the nitramine RDX force field[32], crystalline hexanitrostilbene (HNS, $C_{14}H_6N_6O_{12}$) comprising 32 molecules and using the force field reported by Shan and co-workers[33], and a water box comprising 233 water molecules using the force field of Rahaman et al.[34]. All systems were first equilibrated at the target temperature using the NVT ensemble with a Nosè-Hoover thermostat[35, 36] for 5ps, subsequently the NVE ensemble was applied for another 500 ps to assess the stability of the iEL-SCF method. A time-step of 0.25fs was used in all the simulations apart from the RDX system at 1500K which required a shorter time-step of 0.1fs. The Berendsen thermostat was used to control the dynamics of the auxiliary degrees of freedom. Thermostat temperatures of $10^{-6}\ e^2/fs^2$ and $10^{-5}\ e^2/fs^2$ were used for the $t_{aux}$ and $s_{aux}$ respectively and the thermostat rescaling parameter $\tau$ was set to 0.01ps.

## RESULTS

We evaluate the iEL/SCF approach for charge equilibration using ReaxFF on four systems: liquid water[34], $FeOH_3$[31], hexanitrostilbene (HNS)[37], and nitramine RDX[32] that are standard benchmark systems used in LAMMPS for the ReaxFF force field at room temperature as well as a high temperature for RDX, to evaluate the iEL/SCF improvements. It is important to state at the outset that at least 500 ps is required to determine the underlying behavior of any given CEM solution, unlike previous studies that have characterized algorithms on only 100 fs to 1 ps timescales.

We first determine some measure of a gold standard of convergence in a standard CG-SCF calculation for the CEM solution to determine charges in a standard ReaxFF simulation as implemented in LAMMPS. Figure 1 shows energy conservation under three levels of convergence, $10^{-6}$, $10^{-8}$, and $10^{-12}$ for water and HNS, and the same type of data is reported in Fig. S1 for $FeOH_3$ and RDX. It is clear across the data sets that energy conservation in general is poor using ReaxFF, and although the energy drift is similar once convergence tolerances reach $10^{-8}$ to $10^{-12}$ for the new charges, $q$, the tighter convergence of $10^{-12}$ exhibits more energy drift than $10^{-8}$ in some cases In addition, all plots show that more SCF cycles are required to reach any given level of tolerance for $q_{SCF}^s$ as compared to $q_{SCF}^t$, with the number of total SCF cycles (adding the SCF cycles for $q_{SCF}^s$ and $q_{SCF}^t$) ranging from ~20-25 at $10^{-6}$, ~30-50 at $10^{-8}$, and ~55-120 at $10^{-12}$. Based on this data taken across many different systems, we will consider that convergence at 10⁻⁸ with ~30-40 SCF cycles will be compared in subsequent figures. We note that energy conservation standards differ substantially across the systems, and that more savings will be found if we compare to CG-SCF converged at $10^{-10}$-$10^{-12}$.

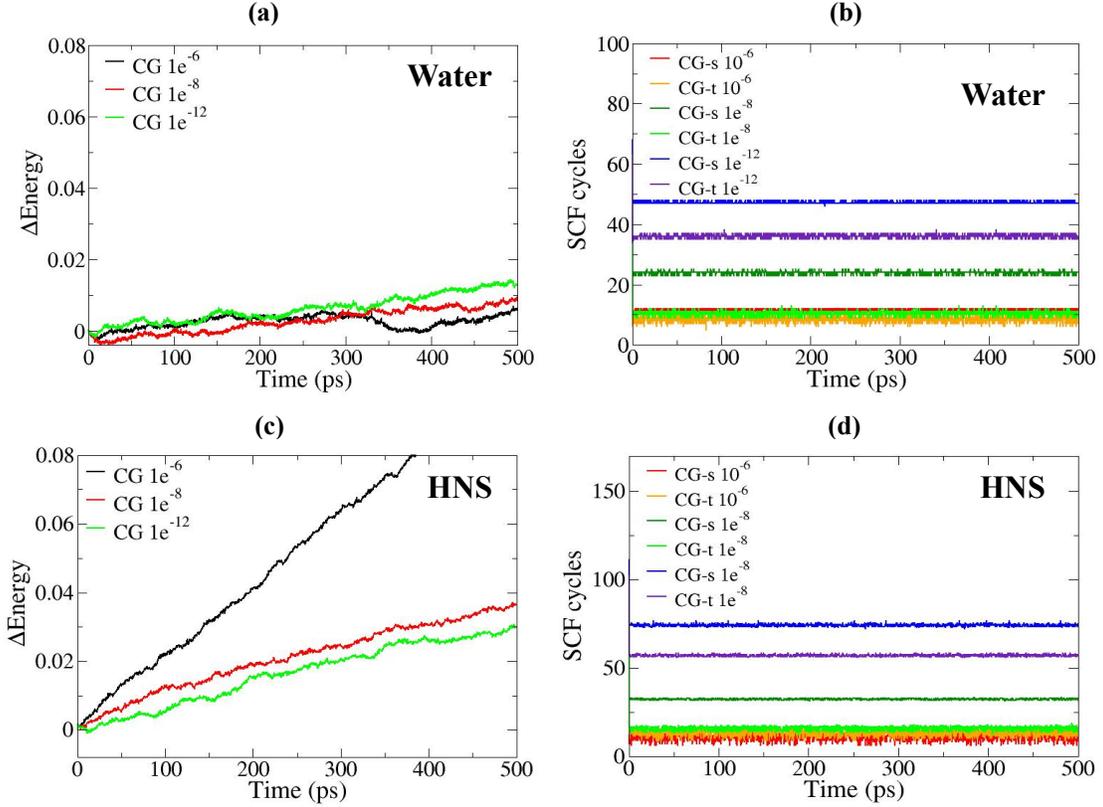

**Figure 1.** *Comparison of energy conservation and number of SCF cycles using the standard CG-SCF solution to charge equilibration.* Comparison of the energy conservation of CG-SCF at $10^{-6}$, $10^{-8}$, $10^{-12}$ level of convergence for (a) water and (c) HNS. Comparison of the number of SCF cycles required to reach convergence at $10^{-6}$, $10^{-8}$, $10^{-12}$ for $q_{SCF}^s$ and $q_{SCF}^t$ for (b) water and (d) HNS. A time step of $\Delta t = 0.25$ was used in all simulations.

The standard XL-BOMD method has been previously applied to the CEM solution for ReaxFF[22], but fixed the number of SCF cycles to a single step. In Figure 2a we compare the energy conservation of this one-iteration XL-BOMD vs. the XL-BOMD method using a loose convergence tolerance of $10^{-2}$, both referenced to the usual CG-SCF at $10^{-8}$. It is evident that the single SCF cycle for $q_{aux}^s$ and $q_{aux}^t$ is highly non-conserving in energy and highly unstable, although of course the cost is reduced by a factor of 30-50 relative to the CG-SCF at $10^{-8}$ convergence. While the XL-BOMD method with loose convergence yields a conserved energy quantity that is nearly as good as the CG-SCF solution, this comes at the computational cost of an unacceptably large number of SCF cycles (Figure 2b) due to corruption of the dynamics of the $q_{aux}^s$ and $q_{aux}^t$ variables from resonance effects that increase the kinetic energy (Figure 2c). The resonance effects are the same as to what has been seen previously using classical polarization[21] or DFT[23], but this impacts the charge equilibration results even more, as it exhibits a much more severe increase in the number of SCF cycles from 10 to more than 45 after only 7.5 ps of molecular dynamics.

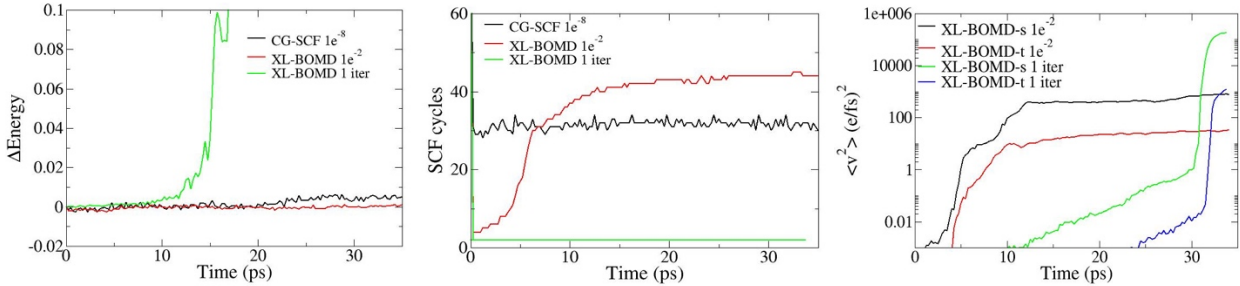

**Figure 2.** *XL-BOMD with no dissipation or thermostats applied to the auxiliary variables.* (a) Comparison of the energy conservation of XL-BOMD using only one iteration, convergence of $10^{-2}$, and CG-SCF at $10^{-8}$ level of convergence. (b) Comparison of the number of SCF cycles required to reach convergence for XL-BOMD at $10^{-2}$ and CG-SCF at $10^{-8}$ vs one iteration. We define the total number of SCF cycles as the sum of the number of SCF cycles for solving $q_{SCF}^s$ and $q_{SCF}^t$. (c) squared velocities $\langle (\dot{q}_{aux}^s)^2 \rangle$ and $\langle (\dot{q}_{aux}^t)^2 \rangle$ for the three approaches. A time step of $\Delta t = 0.25$ was used in all simulations

The benefit of moving to iEL/SCF is that it will control the problems of resonances that plague the standard XL-BOMD approach, through thermostats applied to the auxiliary velocities, $\dot{q}_{aux}^s$ and $\dot{q}_{aux}^t$. Figure 3 shows that the iEL/SCF scheme permits looser convergence of the CEM solution of the new intermediate charges $q_{SCF}^s$ and $q_{SCF}^t$. In Figures 3a-3c we compare the FeOH$_3$, HNS, and RDX systems with the highly encouraging result that there is a significant gain in efficiency using tolerances of $10^{-4}$ for both $q_{SCF}^s$ and $q_{SCF}^t$. In fact energy conservation is as good or even better at this tolerance while only requiring as few as 10 SCF cycles compared to the 35-50 SCF cycles needed by the CG-SCF solver when converging the $q_{SCF}^s$ and $q_{SCF}^t$ to $10^{-8}$. For water the energy conservation is extremely stable and consistent once a tolerance of $10^{-5}$ is reached for $q_{SCF}^s$, and the convergence for $q_{SCF}^t$ can be less tight, on the order of $10^{-3}$ to $10^{-4}$, permitting the number of SCF cycles to drop by half compared to the reference solver (Figure 3d). We note that in the NVT ensemble that these tolerances can be relaxed for water, and thus overall we can recommend that both $q_{SCF}^s$ and $q_{SCF}^t$ can be converged at $10^{-4}$ for any system when using at standard 0.25 fs time step.

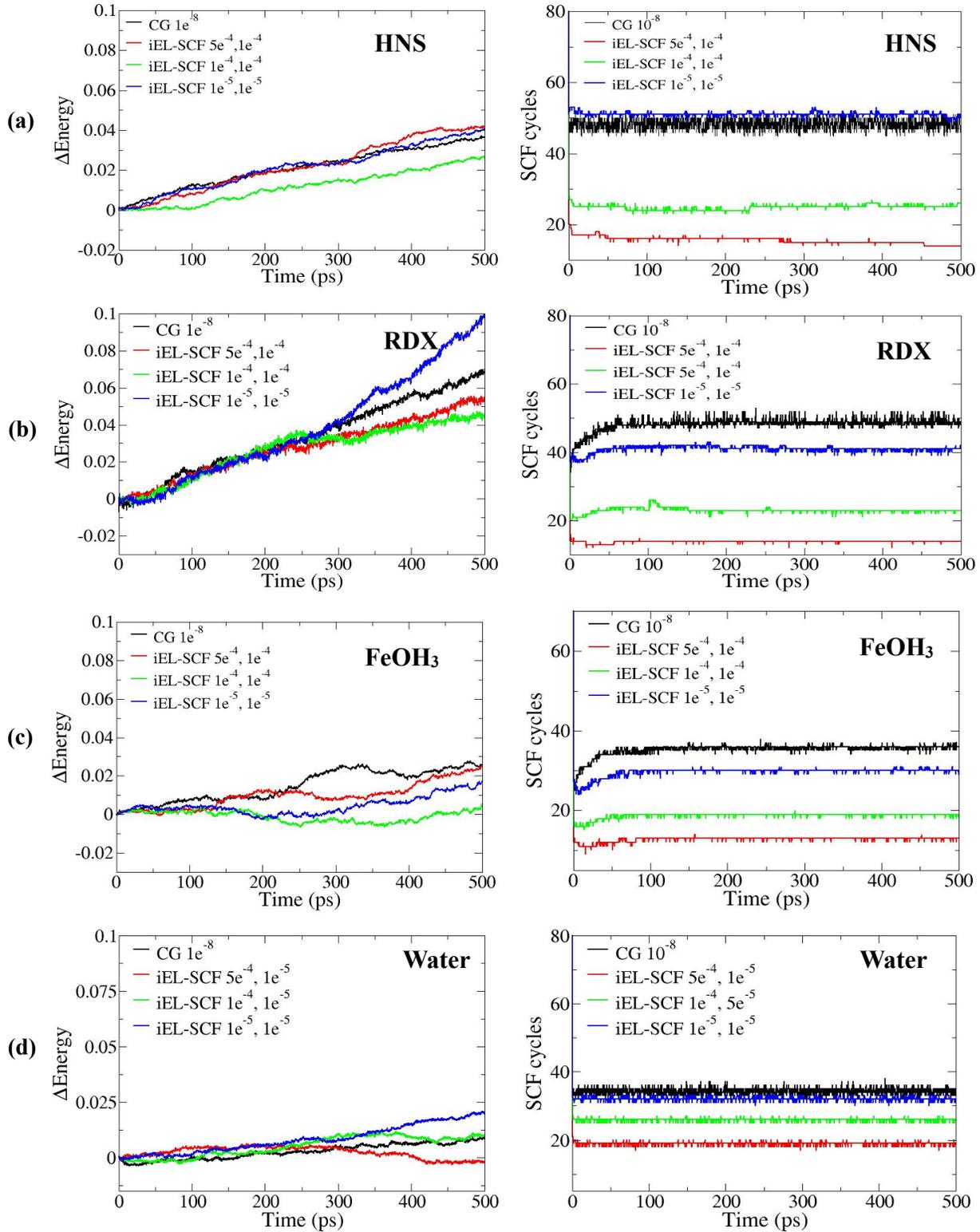

**Figure 3.** *Comparison of energy conservation and number of SCF cycles using the standard CG-SCF solution vs iEL/SCF for charge equilibration.* Comparison of CG-SCF at $10^{-8}$ level of convergence vs. iEL/SCF at convergence of $10^{-5}$ for $q_{SCF}^{s}$ and sweeping through values of $10^{-3}$, $10^{-4}$, and $10^{-5}$ for $q_{SCF}^{t}$ for (a) HNS, (b) RDX, and (c) FeOH$_3$, and (d) water. A time step of $\Delta t = 0.25$ was used in all simulations.

**DISCUSSION AND CONCLUSION**

We note that the general procedure of solving the charge equilibration model using the standard CG-SCF solver shows very poor energy conservation for ReaxFF at the usual time step size of 0.25 fs. Under the three levels of convergence of $10^{-6}$, $10^{-8}$, and $10^{-12}$, it is clear that $10^{-6}$ is not as well converged compared to the more strict tolerances for all four systems based on evident further loss of energy conservation. At the same time there is also no reason to converge as tightly as $10^{-12}$ since there is no gain in energy conservation quality, thereby only increasing the amount of computational work by increasing the number of SCF cycles by a factor of 2.

However further computational efficiency is easily obtained by adopting the iEL-SCF method for solving charge equilibration in LAMMPS as part of the reactive force field ReaxFF. In this case we require the solution to two sets of linear systems and thus require two auxiliary variables for the intermediate charges, both of which serve as an initial guess for the real intermediate charges. In the standard XL-BOMD approach the auxiliary variables are time reversibly integrated together with the atomic degrees of freedom which results in sufficient energy conservation even at loose convergence tolerance. However, as shown before for the case of polarizable force fields[21], the velocities of the auxiliary degrees of freedom become corrupted due to lack of dissipation, which then increases the number of SCF cycles required to reach convergence, thereby defeating its purpose of greater efficiency. We find that the problem of resonances in the case of charge equilibration is much more severe than found for polarizable force fields when using XL-BOMD, resulting in a dramatic increase in the number of SCF cycles even only after a few picoseconds of simulation.

By applying a thermostat for each of the auxiliary charge variables to control their dynamics, the iEL/SCF method is able to achieve stable dynamics and hence reduce the number of SCF cycles to 50-80% of the original CG-SCF solver converged at $10^{-8}$ (more if using $10^{-12}$ CG-SCF as the convergence criteria comparison) without degradation in energy conservation. Because the two sets of linear equations behave differently, we require that the thermostat set point for $q_{aux}^s$ and tolerances for $q_{SCF}^s$ must be controlled more tightly than needed for $q_{aux}^t$ and $q_{SCF}^t$, and suggest that the convergence criteria be set to $10^{-5}$ for the former and either $10^{-3}$ or $10^{-4}$ for the latter, values that work well across all four ReaxFF systems studied. We note that in the NVT ensemble that convergence of $10^{-4}$ will be sufficient, and researchers always have the choice of fine tuning these convergence criteria with shorter runs to determine the best computational performance for their particular ReaxFF force field system.

Recently, the iEL-SCF has been extended to iEL/0-SCF which discards the requirement of an SCF procedure altogether[27-29]. In addition, we have recently formulated a stochastic XL-BOMD procedure[38] that might be usefully combined with iEL/SCF or iEL/0-SCF, as well as exploiting isokinetic schemes with multi-timestepping[30] that may also be useful for increasing the effective time step for CEM. We hope to report on further computational improvements using these methods for CEM in the near future.

**ACKNOWLEDGMENTS.** This work was supported by the U.S. Department of Energy, Office of Science, Office of Advanced Scientific Computing Research, Scientific Discovery through Advanced Computing (SciDAC) program. This research used resources of the National Energy Research Scientific Computing Center, a DOE Office of Science User Facility supported by the Office of Science of the U.S. Department of Energy under Contract No. DE-AC02-05CH11231.